\documentclass[12pt]{article}

\usepackage{cite}

\begin{document}

\title{\bf \boldmath Annotated Bibliography of Some Papers on 
Combining Significances or $p$\,-values}

\author{Robert D. Cousins\thanks{cousins@physics.ucla.edu}
\\ Department of Physics and Astronomy\\ University of California,
Los Angeles, California 90095, USA}
\date{December 20, 2008}
\maketitle

\begin{abstract}
A question that comes up repeatedly is how to combine the results of
two experiments if all that is known is that one experiment had a
$n$-sigma effect and another experiment had a $m$-sigma effect.  This
question is not well-posed: depending on what additional assumptions
are made, the preferred answer is different.  The note lists some of
the more prominent papers on the topic, with some brief comments and
excerpts.
\end{abstract}

\maketitle

\section{Introduction}
\label{intro}

Suppose one experiment sees a 3-sigma effect and another experiment
sees a 4-sigma effect.  What is the combined significance?
Equivalently, given two $p$-values, how can one combine them into one?
Since the question is ill-posed (i.e., more information is needed in
order to specify the best answer), the statistics literature contains
many papers on the topic; a number of them are by prominent
statisticians, including in the 1930's Ronald Fisher and both Karl and
Egon Pearson.  As the existence of this literature may not be
well-known, I have put together this bibliography with some
introductory comments and some brief descriptions and excerpts.  It
should be clear that while these general-purpose methods are useful
for quick estimates (in particular if estimates of relative weights
are available), it is preferable to use additional primary data when
available and to take care to understand the nature of the alternative
hypotheses.

Given a statistic (function of the data) $x$, with (normalized)
probability density function $P(x)$ in the domain $a < x < b$, it is
common in statistics to introduce the {\it probability integral
transform}, letting
\begin{equation}
\label{pit}
y=\int_a^x P(x^\prime)dx^\prime.
\end{equation}
Then the pdf for $y$ is uniform on (0,1), and without loss of
generality many questions about $x$ can be studied in a more
transparent way by considering $y$.  (If $x$ is discrete, there are
complications which are discussed in some of the cited papers.)
Since $1${}$-${}$y$ is also uniform on (0,1), typically some other
consideration (such as the distribution of $x$ under a different $P$)
dictates if one end of the interval is of more interest than the
other.

If $x$ is a test statistic and $P(x)$ its pdf under the null
hypothesis $H_0$, then for one-sided tests of $H_0$ at least a vague
notion of an alternative hypothesis is needed to specify whether
values of $y$ close to 0 or close to 1 should be considered as
evidence against $H_0$.  Then one can identify either $y$ (in the
former case) or $1${}$-${}$y$ (in the latter case) with the $p$-value,
i.e., the smallest value of the significance level $\alpha$ in
Neyman-Pearson hypothesis testing for which $H_0$ would be
rejected. (See \cite{kendall99} for an introduction.)  Here I do not
address the subtleties of Fisherian vs.\ Neyman-Pearson interpretation
of $p$-values, or issues of the utility of $p$-values; I merely
remind the reader that at best, a $p$-value conveys the probability
under $H_0$ of obtaining a value of the test statistic at least as
extreme as that observed, and that it should not be interpreted as the
probability that $H_0$ is true.

Frequently the $p$-value is communicated by specifying the
corresponding number of standard deviations in a one-tailed test of a
Gaussian (normal) variate; i.e., one communicates a $Z$-value (often
called $S$ in high energy physics) given by
\begin{equation}
\label{zdef}
Z = \Phi^{-1}(1-p) = -\Phi^{-1}(p)
\end{equation}
where
\begin{equation}
\Phi(Z) = \frac{1}{\sqrt{2\pi}} \int_{-\infty}^Z \,\exp(-t^2/2)\,dt
\ =\ \frac{1 + {\rm erf}(Z/\sqrt{2})}{2},
\end{equation}
so that
\begin{equation}
\label{eqn:z}
Z = \sqrt{2}\, {\rm erf}^{-1}(1-2p).
\end{equation}
For example, $Z=5$ corresponds to a $p$-value of $2.87 \times
10^{-7}$.

Thus, the question may be asked equivalently as how to combine either
a set of $p$-values or a set of $Z$-values.  While the literature
addresses the problem in both metrics, most of the detailed studies
use the $p$-value, where the two-dimensional version (combining $p_1$
and $p_2$) can be illustrated so transparently: under $H_0$, a scatter
plot of $p_2$ vs $p_1$ uniformly populates the unit square
$(0,1)\otimes(0,1)$, and one desires a function $p(p_1,p_2)$ which is
uniform on (0,1).  Contours of $p$ can be drawn in the $(p_1,p_2)$
square to illustrate each method of combination.

For combining two $p$-values, a general way to construct a combination
method is to first choose some function $f(p_1,p_2)$ which has some
perceived desired properties; then calculate the pdf for $f$ given
that $p_1$ and $p_2$ are uniform on (0,1); and then transform $f$ to
a $p$-value using Eqn.~\ref{pit} with $f$ substituting for $x$.  In
some cases, the function $f$ is by construction already uniform on
(0,1), so that last step is unnecessary.  As for desirable properties
of $f$, the question is so ill-posed that the only property which
is completely general is that of monotonicity, discussed below.  For
example, in the {\it complete} absence of additional information
$f(p_1,p_2) = f(p_2,p_1)$ would seem to be desirable.  However, even
in cases where the details of the combined experiments are not known,
one often knows something about the sample sizes (or more specific
information about the precision of the two experiments), in which case
there is strong motivation to weight the two $p$-values differently.

The ill-posed nature of the problem can be further illustrated by
considering the data of two experiments separately and together.  For
example, suppose two introductory students each make measurements of
current vs.\ voltage across a resistor in order to test the
hypothesis that $I=V/R$, where $R$ is given and fixed.  For
illustration, imagine that all the uncertainty, with normal
distribution, is in the current measurement; that the students make
$N_1$ and $N_2$ measurements, respectively; and that each student then
does a chi-square goodness-of-fit test with d.o.f.'s $N_1$ and $N_2$,
respectively, and computes the $p$-values (from the probability
integral transform of the chi-square pdf), with results $p_1$ and
$p_2$, respectively.  (What is the best goodness-of-fit test is also
ill-posed, but that is another, albeit related, story.)
Alternatively, the data could be pooled and a chi-square
goodness-of-fit test performed with $N_1+N_2$ d.o.f., and a $p$-value
calculated.

In this example, if one has access not to all the data, but only to
the four quantities $p_1$, $p_2$, $N_1$, and $N_2$, then the algorithm
for combine the $p$-values to obtain the pooled answer is clear: use
the inverse of the integral of the chi-square distribution with $N_1$
and $N_2$ d.o.f.'s to recover the two student's chi-squares, add them,
and then use the chisquare integral with $N_1+N_2$ d.o.f. to obtain
$p$.  As with many other algorithms motivated by a specific example,
this algorithm is on the list of general-purpose algorithms which can
be studied in other problems, in which $N_1$ and $N_2$ may be more
artificially chosen to give desired weighting to $p_1$ and $p_2$.  For
this algorithm, the original source commonly cited is
\cite{lancaster61}.

One can also readily see that if nuisance parameters are added,
complications immediately arise.  Thus if $H_0$ is not $I=V/R$ with
$R$ {\it given}, but rather $I=V/R$ where the student is free to fit
for $R$, then immediately we see that d.o.f.'s change and more
information (in particular the best-fit values of $R$ and their
uncertainties) is needed to recover the pooled $p$ (which is still
possible since the uncertainties carry the information of how the two
chi-squares increase as the two fitted values of $R$ are constrained
to an overall best-fit value).

In fact, Lancaster's 1961 method seems to be one of the last general
methods to appear in the literature (and was clearly anticipated), the
others having appeared in the preceding 30 years.  Methods commonly
considered, and names usually associated with them, define the
combined $p$ as follows (for $i=1,N$).

(1) Fisher's method based on the intuitive choice of $f = \prod p_i$.
As the excerpt from his paper below describes, a simple way to
calculate $p$ uses the relation
\begin{equation}
-2\sum \ln p_i  = \chi^2_{2N,p},
\end{equation}
where $\chi^2_{\nu,p}$ denotes the upper $p$ point of the probability
integral of a central chi-squared of $\nu$ degrees of freedom.

(2) Good's generalization of Fisher's method to include weights
$\lambda_i$ so that the test statistic is $Q = \prod p_i^{\lambda_i}$,
with
\begin{equation}
\label{eqn:good}
p = \sum_j \Lambda_j \, Q^{1/\lambda_j}, \ \ 
\Lambda_j =\lambda_j^{N-1} 
\ \prod_{i\ne j}\frac{1}{(\lambda_j-\lambda_i)}.
\end{equation}

(3) Lancaster's generalization of Fisher's method by adding $\chi^2$
    for dof$\ne2$
\begin{equation}
\label{eqn:lancaster}
\sum_i (\chi^2_{\nu_i,p_i})^{-1} 
  = \chi^2_{\nu_{\rm sum},p}; \ \nu_{\rm sum} = \sum_i \nu_i.
\end{equation}

(4) Tippett's method using the smallest $p_i$
\begin{equation}
p = 1 - (1 - (\min \{p_i\}))^N.
\end{equation}

(5) Wilkinson's generalization of Tippet's method, using the $k$th
smallest of the $N$ values of $p_i$.

(6) Stouffer's method adding the inverse normal of the $p_i$'s,
\begin{equation}
\sum \Phi^{-1}(p_i) = \sqrt{N}\Phi^{-1}(p); \ i.e.,\ \ 
Z = \frac{\sum Z_i}{\sqrt{N}}.
\end{equation}

(7) Generalization of Stouffer's method to include weights $w_i$, by
Mosteller and Bush and others, 
\begin{equation}
\label{eqn:mosteller}
Z = \frac{\sum w_i Z_i}{\sqrt{\sum w_i^2}}.
\end{equation}

(8) Lipt\'ak's even more generalized formula with weights which includes
several of the above as special cases: define $Q$ by substituting a
function $\Psi$ for the normal distribution function $\Phi$ in
Eqn.~\ref{zdef}, and proceeding as in the weighted combination of
$Z_i$'s, calculating the distribution of the result and converting to
a $p$-value.

\medskip

It should be clear that {\em any method for combining $p$-values can be used for 
combining $Z$ values, and vice versa}.  E.g., given any two $Z$-values
$Z_1$ and $Z_2$ (normal variates), a combined $Z$-value, also a normal
variant, can be constructed from Eqn.~\ref{zdef} where $p = p(p_1,p_2)$
is obtained by using any $p$-value combination method to
combine $p_1 = 1 - \Phi(Z_1)$ and  $p_2 = 1 - \Phi(Z_2)$.

That this list has grown so long is a testament to the fact that the
question is ill-posed!  When methods differ significantly, say in
combining two $p$-values $p_1$ and $p_2$, the difference is typically
in how they rank the combination of two similar $p$-values compared to
the combination of a high one and a low one.  Which ranking is
preferred depends of course on which parts of the unit square (nearer
the axes or nearer the center) the alternative hypotheses tend to
populate.

The remainder of this note mentions a number of
papers.  Section \ref{primary} lists notable primary papers.  Section
\ref{briefreviews} lists several reviews which compare some of the
above methods.  Section \ref{applications} list some papers with
applications in the life, physical, and social sciences.  In most
cases, I retain the original author's notation, which corresponds to
the above in a transparent way.  Most of the papers are readily
available on the web, in particular at www.jstor.org, which however
requires an institutional license.

\section{Notable Primary Sources}
\label{primary}

Sir Ronald Fisher's book, {\em Statistical Methods for Research
Workers}, first appeared in 1925 and has been enormously influential
through its many editions.  As cited by Karl Pearson \cite{pearson33},
the method for combining significance levels appears to have been
introduced in the 4th edition of 1932.  The 14th Edition of 1970 reads
\cite{fisher70},

``When a number of quite independent tests of significance have been
made, it sometimes happens that although few or none can be claimed
individually as significant, yet the aggregate gives an impression
that the probabilities are on the whole lower than would often have
been obtained by chance.  It is sometimes desired, taking account only
of these probabilities, and not of the detailed composition of the
data from which they are derived, which may be of very different
kinds, to obtain a single test of the significance of the aggregate,
based on the product of the probabilities individually observed.

``The circumstance that the sum of a number of values of $\chi^2$ is
itself distributed in the $\chi^2$ distribution with the appropriate
number of degrees of freedom, may be made the basis of such a test.
For in the particular case when $n=2$, the natural logarithm of the
probability is equal to $-{1\over2}\chi^2$.  If therefore we take the
natural logarithm of a probability, change it sign and double it, we
have the equivalent value of $\chi^2$ for 2 degrees of freedom.  Any
number of such values may be added together, to give a composite
test\dots ''

As emphasized by several authors, Fisher's principle for combining
$p$'s is the last line of the first paragraph quoted; the second
paragraph is a technical implementation equivalent to performing the
probability integral transformation of the product.

Karl Pearson \cite{pearson33} independently proposed the same test and
a variant using $1-p$ instead of $p$, the latter of which is sometimes
called Pearson's method even though most of his paper is on the same
method as Fisher, with more voluminous discussion.  He says in a
``Note added'' that, ``After this paper had been set up Dr Egon
S. Pearson drew my attention to \dots R.A. Fisher's \dots''.  (Egon,
the Pearson of Neyman-Pearson Lemma fame, was the son of Karl, the
Pearson of Pearson's chi-square.)

E.S. Pearson \cite{pearson38} briefly reviews the probability integral
transformation, ``which seems likely to be one of the most fruitful
conceptions introduced into statistical theory in the last few
years'', and looks at examples of the early methods.  He notes the
``difference in character'' between common alternative hypotheses
(goodness of fit problem) and that of different alternative hypotheses
for each of the $p$'s.  Later he \cite{pearson50} considered
extensions to discrete distributions such as binomial and Poisson.

Tippett \cite{tippett31} 
was apparently the first to suggest rejecting $H_0$ at
significance level $\alpha$ when {\em any} of $p_1,\dots,p_k$ is less
than or equal to $1-(1-\alpha)^{1/k}$.  I.e., one uses only the
smallest $p$ and corrects for the effect of having $k$ tries to attain
it.  Tippett's method was generalized in a short note by Wilkinson
\cite{wilkinson51} to the case where one observes $n$ or more
significant statistics in a set of $N$.

In a landmark sociological study using data from exit interviews of
American soldiers, Stouffer et al.~\cite{stouffer49} specified the
method which became known as ``Stouffer's method'' in an obscure
footnote (!), adding three $Z$-values (obtained from a Gaussian
approximation to binomial data) and dividing by $\sqrt{3}$. (The
question was whether men with a better attitude had a better chance of
promotion, and a positive effect at significance level 5\% was found.)

Birnbaum \cite{birnbaum54} evaluated several methods (Fisher, Pearson,
Tippett, Wilkinson) in terms of generally desirable properties such as
monotonicity and admissibility.  ``A test is admissible if there is no
other test with the same significance level which, without ever being
less sensitive to possible alternative hypotheses, is more sensitive
to at least one alternative.''  He states ``Condition 1: If $H_0$ is
rejected for any given set of $u_i$'s, then it will also be rejected
for all sets of $u_i^\ast$'s such that $u_i^\ast \le u_i$ for each
$i$.'' Then, ``\dots the question is whether any further reasonable
criterion can be imposed to narrow still further the class of methods
from which we must choose.  The answer is no\dots''. ``These
considerations prove that to find useful bases for choosing methods of
combination, we must consider further the particular kinds of tests to
be combined\dots''.  For most of the problems he considers, ``Fisher's
method appears to have somewhat more uniform sensitivity to the
alternatives of interest\dots''.  Birnbaum also emphasizes that the
alternative hypothesis to $H_0$ depends on the experimental situation
and in particular there are two classes, which he calls $H_A$: All of
the $u_i$'s have the same (unknown) non-uniform, non-increasing
density $g(u)$; and $H_B$: One or more of the $u_i$'s have (unknown)
non-uniform, non-increasing densities $g_i(u_i)$.

Good \cite{good55} generalized Fisher's product-of-$p$s method in
order to accommodate different positive weights for the results to be
combined.  He inserted the weights as different exponents for each
$p$-value in the product, and derived the distribution of the
resulting test statistic (assuming unequal weights).

Yates \cite{yates55}, in a paper devoted to issues of combining
(discrete) data from $2\times2$ tables, begins by saying that the
method of maximum likelihood is preferable, but for quick, possibly
preliminary tests, combining via tests such as Fisher's test with
$n=2$ (which he takes as the usual test without attribution) ``may be
regarded as adequate''. Under ``Variants of the test'', he writes:
``The use of values of $\chi^2$ for 2 d.f. for the combination of
probabilities is to a certain extent arbitrary.  It has the
convenience that the values are easily calculated, and the use of a
function of the product of the probabilities has a certain intuitive
appeal, but the method would work equally well with other basic
numbers of degrees of freedom.  If, for instance, the values of
$\chi^2$ for 1 d.f. corresponding to the $P$'s are summed then in the
absence of association the sum will be distributed as $\chi^2$ for $k$
d.f.''.  He also considers the test that appears to be Stouffer's test
in this context, describing the signed, normalized deviations as
``\dots normal deviates with zero mean and unit standard deviation,
and their sum is therefore a normal deviate with a standard deviation
of $\sqrt{k}$'', citing as an example Cochran \cite{cochran54}.
(\cite{cochran54} says that the test criterion, $\sum X/\sqrt{g}$
using standard normal tables ``has much to commend it if the total
$N$'s of the individual tables do not differ greatly (say by more than
a ratio of 2 to 1) and if the $p$'s are all in the range 20\%-80\%.'')
After considering these variants applied to some examples, Yates does
not see much difference and recommends Fisher's test ``on historical
grounds and because of its simplicity and intuitive appeal'' if one is
given $p$-values, but Cochran's combination if one is given the unit
variates (as he sees little point in transforming them, as these are
only quick approximate calculations).  He summarizes
unenthusiastically: ``Reasons are given for believing that combination
of probabilities tests are not likely to be very efficient\dots''.

In 1958, Lipt\'ak \cite{liptak58} published a very useful overview,
unifying and generalizing the theory of the various methods on the
market, and elucidating the criteria for a method to be admissible.
(Some of this is similar to Birnbaum's work, of which Lipt\'ak seems to
have been unaware.)  Lipt\'ak's paper, published in a Hungarian journal,
unfortunately was overlooked by some subsequent authors, and even
today is not available online. He defines the combination problem as
that in which ``either the null-hypothesis is true in each experiment
or the alternative one is valid in each case''.  He introduces an
``averaging'' function $\chi$, a strictly increasing and continuous
function with domain (0,1); in practice $\chi$ can be thought of as
the {\it inverse} of the probability integral transformation in
Eqn.~\ref{pit}, i.e., it takes a $p$-value back to the metric used in
some pdf $P$. He further introduces weights $\lambda_i$ for each
$p_i$; the combined test statistic is then $\sum \lambda_i\chi(p_i)$,
the pdf of which can be calculated and transformed to a $p$-value
using again Eqn.~\ref{pit} in the forward direction.  (Thus his term
``averaging function'' refers not to the weights, but rather to the
function of $p$ which is to be averaged.)  Lipt\'ak restricted his paper
to the case where the $\chi$'s are all the same (unlike Lancaster).
The ill-posed nature of the combination problem is apparent, since
(with reasonable assumptions) he shows that for every choice of $\chi$
and weights $\{\lambda_i\}$, there exists a hypothesis testing problem
(i.e., an alternative hypothesis) for which that choice is the optimal
solution (in terms of Type II error probability) of the combination problem.
He also shows that his assumptions (in particular monotonicity) are
consistent with Bayesian solutions to the combination problem, noting
``The importance of this theorem is clear from the fact that the class
of all Bayes solutions in a relatively wide and typical class of
hypothesis testing problems are `complete', i.e., for every test there
can be given a Bayes solution which is at least as good as this test''
(citing A.~Wald's book on statistical decision functions).   
Lipt\'ak shows
that methods such as those of Fisher and Stouffer correspond to
different choices of $\chi$ and $\lambda_i$.  As an omnibus test he
advocates the weighted version of Stouffer's test
(Eqn.~\ref{eqn:mosteller}), in which the the weights ``should be
chosen as to express the efficiencies\dots of the tests used in the
individual experiments.''  In a common simple case, this leads to
$\lambda_i = \sqrt n_i$, where $n_i$ is the number of observations in
the $i$th experiment, as noted by others as well.

As referred to above, in 1961 Lancaster \cite{lancaster61}
(apparently unaware of Lipt\'ak's paper)
generalized Fisher's method: ``Let us suppose that $P_i$ of the $i$th
experiment is transformed to the scale of $v_i = \chi^2$ with $s_i$
degrees of freedom, and let the simple sum be formed, $V^\ast =
\sum_{i=1}^N v_i$, then $V^\ast$ is $\chi^2$ with $\sum s_i$ degrees
of freedom.''  He then describes how $V^\ast$ can be evaluated using
asymptotic properties of the $\chi^2$ distributions (whereas nowadays
one can numerically calculate the tail property, as indicated in
Eqn.~\ref{eqn:lancaster}).  He notes that weights can be introduced as
follows ``\dots if the weights of the different experiments are
different, the variation in the degrees of freedom\dots will give
weights to the experiments proportional to the square root degrees of
freedom\dots '' Remarkably, Lancaster goes on to say, ``On the other
hand, it will usually be simpler to obtain standardized normal
variables and sum them.  Weighting is then easily introduced as
multipliers\dots'', thus describing the method identical to Stouffer's
method with weights (Eqn.~\ref{eqn:mosteller}) while identifying it
only with Yates \cite{yates55}.  Lancaster concludes that it does not matter
greatly whether one uses Fisher's method, the method of normal
variates, or his own method, while noting that the normal variate
method is computationally easier when their are weights.  (As a
postscript, in 1967 Lancaster wrote, while discussing the various
methods in answer to a query \cite{lancaster67}, that Fisher thought
it would be improper to generalize the transformation to $n\ne2$ dof.)

Oosterhoff \cite{oosterhoff69} wrote a monograph on the combination of
one-sided tests, including a historical introduction, various
theorems, and some graphs of acceptance regions.  (This followed
earlier work by Zwet and Oosterhoff \cite{zwet67}.)  Sprott
\cite{sprott71} reviewed the book, describing the ``interesting
historical survey'' but otherwise finding it of narrow interest and
that ``it would appear to have limited value to a practising
statistician involved with practical problems.''

Berk and Cohen \cite{berk79}, consider a criterion of optimality known
as Bahadur relative efficiency, and categorize methods as
asymptotically Bahadur optimal (ABO) or not.  Fisher's method is ABO
\cite{littell71,littell73}, but there are many other ABO unweighted
methods.  \cite{berk79} is particularly interesting because it
considers weighted methods as well.  It concludes that Lancaster's
method \cite{lancaster61} is ABO but Good's method \cite{good55} is
not.  They describe Lancaster's method in terms of the $\Gamma$
function rather than the related chi-square distribution, defining
$W_i = \{\Gamma_{\alpha_i}^{-1}(1-L(T_i, n_i))\}/n$, where
$\Gamma_{\alpha_i}$ is the gamma cumulative distribution function,
with parameters $\alpha_i$ and ${1\over2}$. (The $L$'s are the
$p$-values.)  ``There is complete flexibility in the choice of the
$\alpha_i$'s, which play the role of weights \dots The statistic
$W_\Gamma = \sum W_i$ is such that $nW_\Gamma$ has a $\Gamma(\sum
\alpha_i, {1\over2})$ distribution, so that critical values are
readily attainable from chi-squared tables if $\sum\alpha_i$ is an
integer.''

Wright \cite{wright92} discusses, with numerous references, the
problem of adjusting the $p$-value of an individual test, when taken
in the context of other tests.  While this seems to be closely related
to the problem of combining $p$-values, the literature appears to be
disjoint from that in the rest of this bibliography, and I have not
pursued it.

For combining tests of correlation coefficients, Han \cite{han89}
proposed a test based on a weighted linear combination of Fisher $z$
transformations of the $p$-value.  N.I. Fisher \cite{fisher90}
commented that the Lancaster \cite{lancaster61} generalization of
Fisher's test was of interest to try as well, and in the reply to
comment, Han says that Lancaster's method was better than the
unweighted Fisher test, but had smaller power than Han's test in most
cases.

For the special case of ``balanced incomplete block design'' Mathew et
al. \cite{mathew93} describe a combination procedure which they say
outperforms Fisher's method (which they suggest is likely to be
inadmissible).  They also emphasize the distinction between common and
separate alternative hypotheses.

Goutis et al.\ \cite{goutis96} attempt to state formally their
``axioms'' which a $p$-value combination scheme should satisfy, based
partially on comparison with a Bayesian model.  They cite
\cite{birnbaum54} but make the point that decision theory may not be a
reliable guide.  The discussion points to the need for more
information about the experiments than just the $p$-values,
noting that combining combinations of $p$-values is problematic.
``We can say that evidential measures based on combining rules of
Fisher and Tippett seem to perform reasonably\dots''.  Regarding the
decision theoretic approach and their axiomatic approach, ``\dots we
are unable to reconcile the two approaches\dots''.

\section{Reviews}
\label{briefreviews}

In comparing methods, reviews necessarily consider some classes of
alternatives to $H_0$, and hence the conclusions can vary, or even
contradict each other, depending on the alternatives chosen.  

Rosenthal \cite{rosenthal78} surveys nine methods used in psychology,
and says that ``the seminal work of Mosteller and Bush
\cite{mosteller54} is especially recommended''.  He finds limitations
with the Fisher method, and concludes: ``There is no best method under
all conditions \cite{birnbaum54}, but the one that seems most
serviceable under the largest range of conditions is the method of
adding $Z$s, with or without weighting'' (i.e., Stouffer's method,
with or without modification by Mosteller and Bush).

The same year, Koziol and Perlman \cite{koziol78} considered the
non-central chi-square problem, including the sum of chi-squares
statistic with different d.o.f., emphasizing that different
combination methods have advantages depending on the alternative
hypothesis. One of their conclusions was, ``It is difficult to
recommend the inverse normal procedure in any circumstance''.  This
is remarkable in view of Rosenthal's conclusion and the apparent 
popularity of Stouffer's method in psychology.

Loughin \cite{loughin04} studies six methods.  He cites Lipt\'ak
\cite{liptak58} for the unweighted $z$ method instead of the more
usual Stouffer citation, and cites Mosteller and Bush only to say that
he (Loughin) does not consider weights in this paper.  He has some
nice comparisons in the unit square, and notes that all the methods
considered satisfy a property of {\em monotonicity}: ``rejecting $H_0$
for $\{p_1,\dots,p_k\}$ implies rejecting $H_0$ for all
$\{p_1^\ast,\dots,p_k^\ast\}$ such that $p_i^\ast \le p_i$,
$i=1,\dots,k$.  Holding everything else constant, greater evidence
against $H_{0i}$ implies greater evidence against $H_0$.''

Marden, in two papers which begin with a brief review, studies the
performance of many of the above tests when applied to non-central
chi-squared tests or $F$ tests \cite{marden82}; and non-central $t$ or
normal mean tests \cite{marden85}.  The inverse normal procedure
(Stouffer's test) is found to be inadmissible in both papers.

In their book on of meta-analysis (the modern term for combining
results from different experiments or trials), Hedges and Olkin
\cite{hedges85} devote Chapter 3 to methods for combining tests,
reviewing all of the methods described above.  They conclude, ``It
seems that Fisher's test is perhaps the best one to use if there is no
indication of particular alternatives''.

\section{Some papers with Applications and Discussion}
\label{applications}

In high energy physics (HEP), the first and second editions of the
popular text \cite{eadie71},\cite{james06} describe Fisher's method.
Cousins \cite{Cousins:1993gs} discusses Fisher's method and
generalizations to $n\ne2$, in particular to $n=1$ by G. Irwin, but
was apparently aware of neither \cite{lancaster61} nor Fisher's
insistence \cite{lancaster67} on $n=2$ due to the
product-of-$p$-values derivation.  Recently in HEP, Bityukov et
al. \cite{bityukov06} advocate a method which is essentially
Stouffer's method (without the generalization
\cite{mosteller54},\cite{liptak58}\cite{lancaster61} to include weights).

In a more substantive HEP paper, Janot and Le Diberder
\cite{Janot:1998fz} discuss various optimality and reasonableness
considerations and propose a method with a weighted combination scheme
which can be identified as that of Good \cite{good55}.  (The formula
for the combined $p$-value (Eqn.~\ref{eqn:good} above) is their
Eqn.~(38) and Eqn.~(1) of \cite{good55}.)  They discuss in detail the
choice of optimal weights, as well as extensions to the discrete case,
in particular low-statistics Poisson variates encountered in frontier
HEP experiments, with consideration of reducible and irreducible
background.

In astrophysics, Afonso et al. \cite{afonso03} use Lipt\'ak's
\cite{liptak58} weighted generalization of Stouffer's method in a
search for gravitational lensing events, using as weights the expected
number of events of each experiment for each lens mass.

In evolutionary biology, Whitlock \cite{whitlock05} compares Fisher's
method and the weighted Stouffer method of Mosteller and Bush
\cite{mosteller54}, and concludes that the latter is better; this is
perhaps not surprising since it uses additional information.  He notes
that Stouffer's method was originally ``given in a footnote on p.\ 45
of their sociological work on the Army \dots This must be one of the
most obscure origins of a statistical method in the literature''.

Zaykin et al. \cite{zaykin02}, in gene studies, review possibilities
and advocate a variant of Fisher's method in which only $p$-values
below some truncation value are used.  Among the advantages, they
state, ``Experience shows that the ordinary Fisher product test loses
power in cases where there are a few large $p$-values.''  They study
power in cases relevant for them. Neuh\"auser and Bretz
\cite{neuhauser05} do some studies of the truncation.

In molecular biology, Koziol \cite{koziol96} explores the relationship
between the weighted methods of Good \cite{good55} and Lipt\'ak
\cite{liptak58}.  Koziol and Tuckwell \cite{koziol99} consider as well
these tests from a Bayesian point of view.  ``Fisher's well-known
combination procedure\dots is found to be a Bayes test in this setting
with a noninformative prior.  Good's weighted version of Fisher's
procedure is shown to be an excellent approximation to other Bayes
tests.''

Psychology professors Darlington and Hayes \cite{darlington00}
consider the Stouffer $z$ method to be the best-known method, and
propose extensions to it and other methods.  They note, ``As a general
rule, the Stouffer method yields a numerically smaller (thus, more
significant) pooled $p$ than the Fisher method if the entering $p$s
are fairly similar in value, whereas that Fisher method yields a more
significant result if the entering $p$s vary widely.''  They advocate
the method they call Stouffer-max, which considers only the $s$ most
significant results (out of $k$).

In economics, Wallis \cite{wallis42} discusses Fisher's solution,
emphasizing forcefully the derivation based on product of
probabilities, lamenting that it is misunderstood that $\chi^2$ with
$n=2$ was chosen for convenience: ``This reasoning is entirely
fallacious\dots that transformation, when used, is valid only for two
degrees of freedom.''  But having said that, he proceeds to
investigate if the product of the probabilities is indeed the proper
criterion of joint significance, and does so using the Neyman-Pearson
Lemma for comparison.  He concludes the Fisher's test probably
excludes a region ``not radically different from the ideal region'' in
the kind of problems ``most likely to be found practical work''.  He
goes on to consider the discrete case.  As another indication of the
galaxy of people who have thought about this problem, it is
interesting that Wallis is ``indebted to Milton Friedman for
collaborating in the investigations'' of one section of the paper and
for ``many stimulating discussions''.

Louv and Littell \cite{louv86} compare six of the usual methods in the
case of tests on a binomial parameter, with detailed discussion of the
circumstances in which each has advantages or disadvantages, also
making the distinction of whether or not the alternative is common to
all $p_i$'s.

Westberg \cite{westberg85} compares Fisher's method and Tippett's
method, and discusses when she recommends each.  ``The selection of a
combination procedure is difficult.  You should consider which is the
most interesting deviation from $H_0$ if the null hypothesis is not
true, and also the number of interesting hypotheses that are false.''
``Bahadur's efficiency is an asymptotic concept \dots the asymptotic
result does not actually carry over to finite samples.''

\medskip
{\bf Acknowledgments} I thank Louis Lyons for stimulating,
educational discussions, and Luc Demortier for helpful comments on the
manuscript and for providing additional references.  This work was
partially supported by the U.S. Dept.\ of Energy and by the National
Science Foundation.

\bibliography{combine_bibtex}{}
\bibliographystyle{apalike}
\end{document}